# 220 fs Er-Yb:glass laser mode-locked by a broadband low-loss Si/Ge saturable absorber

F. J. Grawert, J. T. Gopinath, F. Ö. Ilday, H. Shen, E. P. Ippen and F. X. Kärtner

Department of Electrical Engineering and Computer Science and Research Laboratory of Electronics, Massachusetts Institute of Technology, Cambridge, MA 02139

S. Akiyama, J. Liu, K. Wada, L. C. Kimerling

Department of Materials Science and Engineering and Material Processing Center, Massachusetts Institute of Technology, Cambridge, MA 02139

**We demonstrate femtosecond performance of an ultra-broadband high-index-contrast saturable Bragg reflector consisting of a silicon/silicon-dioxide/germanium structure that is fully compatible with CMOS processing. This device offers a reflectivity bandwidth of over 700 nm and sub-picosecond recovery time of the saturable loss. It is used to achieve mode-locking of an Er-Yb:glass laser centered at 1540 nm, generating 220 fs pulses, with the broadest output spectrum to date.**

Passive mode-locking with a saturable Bragg reflector (SBR) is a powerful method to generate a steady stream of pico- or femtosecond pulses from a laser [1], [2]. To date, SBRs have been fabricated both as bulk and quantum well devices from III-V compound semiconductor materials which are not compatible with the silicon material platform. In contrast, the silicon/germanium SBR (Si/Ge-SBR) demonstrated here consists of a CMOS-compatible silicon/silicon-dioxide (Si/SiO$_2$) Bragg reflector and a germanium saturable absorber layer (Figure 1a). Due to the high refractive index contrast ($n_{SiO_2}$=1.45 and $n_{Si}$=3.5), only six layer pairs in the Bragg mirror are sufficient to achieve a maximum reflectivity of 99.8%. On top of the Si/SiO$_2$ Bragg stack, a germanium saturable absorber layer is embedded in a silicon layer of 3λ/4 optical thickness at the center wavelength of 1400 nm. It resides at a peak of the standing wave pattern of the electric field to maximize absorption and minimize saturation intensity (Figure 1a).

The manufacturing of the SBR was aimed at achieving both a high reflectivity of the Bragg mirror and a sufficiently-strong nonlinear response of the absorber layer for mode-locking an Er-Yb:glass laser. The 6-pair reflector was fabricated by repeating poly-Si deposition and thermal oxidation using a silicon-on-insulator (SOI) wafer as the starting material (Figure 1b, step 1). Then, the Bragg mirror was bonded with a new silicon substrate (Figure 1b, step 2). The silicon handle and buried oxide of the SOI-wafer were chemically etched to expose the crystalline silicon layer of the SOI-wafer for successive germanium epitaxial growth (Figure 1b, step 3). A 40 nm thick germanium saturable absorber layer was deposited by the UHV-CVD technique developed by Luan *et al*. [3] (Figure 1b, step 4). Finally, a thin oxide was grown on germanium as a passivation layer and a poly-Si cap layer was deposited. The manufacturing process including bonding of a new substrate and removal of the original one results in the reversal of the Si/SiO$_2$ layer sequence of the reflector with respect to layer growth. Thus, the layers of lowest surface roughness – those that have been grown first – ending up topmost in the mirror, exposed to the highest electric field strength, while the rougher layers, grown last, are buried deep in the mirror (Figure 1a). This reversal of the layer sequence significantly reduces surface roughness and scattering losses, leading to an unprecedented 99%-reflectivity bandwidth of 700 nm (Figure 2b). Furthermore, substrate reversal terminates the reflector with the crystalline silicon layer of the SOI-wafer rather than with polycrystalline material deposited on a Bragg mirror. It is the crystalline nature of this silicon layer that preserves the optical absorption properties of the germanium layer to achieve saturable absorption in the silicon / germanium material system at 1550 nm.

The nonlinear response of the device was characterized in a series of pump-probe measurements with 150 fs pulses centered at 1540 nm from an optical parametric oscillator. For low to medium fluence values (*e.g.* 40 µJ/cm$^2$), the germanium layer acts as a fast saturable absorber with up to 0.13% of modulation depth (Figure 2a). We observe sub-picosecond recovery of the bleaching, with the temporal resolution of our measurement being limited by the pump and probe pulse durations. We attribute the fast relaxation process to intervalley scattering within the conduction band. In contrast, for high fluences (*e.g.*, 300 µJ/cm$^2$), carriers generated by two-photon absorption (TPA) induce free carrier absorption (FCA) and turn the germanium layer to an inverse saturable absorber. The strong inverse saturable absorption of the Si/Ge-SBR is due to the TPA in the germanium layer ($\beta_{Ge,1500nm}$ = 300 cm/GW) which is much greater than that of silicon or gallium arsenide. The observed behavior leads to dual functionality of the Si/Ge-SBR in a mode-locked laser: (i) the fast recovery enables ultrashort pulse generation, (ii) onset of inverse saturable absorption at high fluences helps stabilize high repetition rate lasers against Q-switching by limiting the maximum intracavity power [4], [5] . This instability has prevented successful mode-locked operation of lasers with long upper state lifetimes and high repetition rates until recently [6], [7] and is a major obstacle for compact laser integration. The thin germanium layer grown on silicon is compressively strained, leading to a shift of the bandgap by 38 nm to shorter wavelengths. As a result, absorption of the germanium sets in at 1580 nm (Figure 2b), leading to a total loss of 0.3%, and a nonsaturable loss of 0.17% at the laser wavelength, as determined by comparison of the intracavity power obtained with the Si/Ge-SBR with different output couplers. In addition, a transmission electron micrograph revealed a small error in the SOI-layer thickness, placing the Ge-absorber layer slightly off the peak of the electric field. A larger modulation depth of the SBR can be expected from a precise positioning of the Ge layer at a field maximum, as well as by the use of a thicker layer. From pump probe measurements we estimate the saturation fluence to be about 30 $\mu J/cm^2$.

The fast recovery time of the Si/Ge-SBR combined with its high reflectivity leads to superior spectral width and pulse duration of the bulk Er-Yb:glass laser mode-locked with this device. A phosphate glass Kigre QX/Er served as the gain medium in the laser. It is flat-Brewster polished, placed at one end of the four-element laser cavity (Figure 3a), and its flat side serves as an output coupler with 99.8% reflectivity. The laser is pumped with a 450 mW fiber-coupled diode laser (Bookham, type G07). The overall intracavity loss is minimized with highly-reflecting mirrors, low output coupling and a highly-reflecting Si/Ge-SBR, leading to an average intracavity power of 8.7 W. The laser is operated with a highly-saturated gain resulting in a flat gain profile to support a broad optical spectrum [8]. We obtain an optical spectrum centered at 1550 nm with a full-width half-maximum (FWHM) bandwidth of 11 nm and covering the entire C-band of optical communications at approximately –10 dB level (Figure 3b). After dechirping the pulses extracavity with 1.0 m of single mode fiber (Corning SMF-28), the inferred pulse duration from the intensity autocorrelation is 212 fs (Figure 3d). Phase retrieval with an iterative algorithm [9] reveals a pulse width of 220 fs, which is 10% larger than the transform-limited value, obtained from the zero-phase Fourier transform of the power spectrum. To our knowledge, these are the shortest pulses generated from a bulk Er/Yb:glass laser to date [10], and about an order of magnitude shorter than those obtained solely from mode-locking of an Er/Yb:glass laser with an SBR [11]. The laser operates at a 169 MHz repetition rate with a clean RF spectrum and a noise floor more than 70 dB below the signal level (Figure 3c). No Q-switching behavior was observed regardless of pump power level despite the long upper-state lifetime and the small emission cross section of the gain medium. We attribute the high stability against Q-switching to the inverse saturable absorption in the Si/Ge-SBR at high fluence.

In conclusion, we have demonstrated a silicon/germanium saturable Bragg reflector fabricated with a CMOS compatible process. Its nonlinear response has been characterized by femtosecond pump-probe measurements showing fast saturable absorber behaviour on a femtosecond scale and strong inverse saturable absorption at large fluence values. The device has been used to attain self-starting operation of a passively mode-locked Er-Yb:glass laser with an optical spectrum covering the entire C-band of optical communications. The larger modulation depth of the present device at shorter wavelengths with the broad bandwidth of the $Si/SiO_2$ backmirror can be utilized in mode-locking other laser systems, such as Cr:Forsterite and $Cr^{4+}$:YAG, which support few-cycle laser pulses. The development of a saturable absorber in the silicon material platform paves the way to the construction of chip-scale mode-locked lasers in the near future. We envision such lasers to become compact, low-noise and inexpensive light sources, that allow for new applications in optical communications [12], high-speed optical sampling [13], on-chip clocks [14] and low noise microwave oscillators [15].

This research was supported by NSF under grants ECS-0322740, ONR-N00014-02-1-0717 and AFOSR FA9550-04-1-0011. The authors thank Bookham Technology Plc. for providing a 980 nm pump diode.

**References:**

1. G. Steinmeyer, D. H. Sutter, L. Gallman, N. Matuschek, U. Keller, *Science* **286**, 1507-1511 (1999)

2. S. Tsuda, W. H. Knox, S. T. Cundiff, W. Y. Jan, J. E. Cunningham, *IEEE J. sel. Topics in Quantum Electronics* **2**, 454-464 (1996)

3. H.-C. Luan, D. R. Lim, K. K. Lee, K. M. Chen, J. G. Sandland, K. Wada, L. C. Kimerling, *Appl. Phys. Lett.* **75**, 2909-2911 (1997)

4. E. R. Thoen, E. M. Koontz, M. Joschko, P. Langlois, T. R. Schibli, F. X. Kärtner, E. P. Ippen, L. A. Kolodziejski, *Appl. Phys. Lett.* **74**, 3927 -- 3929 (1999)

5. T. R. Schibli, E. R. Thoen, F. X. Kärtner, E. P. Ippen, *Appl. Phys. B* **70**, 41 -- 49 (2000)

6. P. Langlois, M. Joschko, E. R. Thoen, E. M. Koontz, F. X. Kärtner, E. P. Ippen, L. A. Kolodziejski, *Appl. Phys. Lett.* **75**, 3841 - 3483 (1999)

7. S. C. Zeller, F. Krausz, G. J. Spuehler, R. Paschotta, M. Golling, D. G. Ebling, K. J. Weingarten, U. Keller, *Electron. Lett.* **40**, 875-876 (2004)

8. C. Hoenninger, R. Paschotta, M. Graf, F. Morier-Genoud, G. Zhang, M. Moser, S. Biswal, J. Nees, A. Braun, G. A. Mourou, I. Johannsen, A. Giesen, W. Seeber, U. Keller, *Appl. Phys. B* **69**, 3-17 (1999)

9. J. W. Nicholson, J. Jasapara, W. Rudolph, *Opt. Lett.* **24**, 1774-1776 (1999)

10. G. Wasik, F. W. Helbing, F. Koenig, A. Sizmann, G. Leuchs, CLEO proceedings, CMA4 2001

11. G. J. Spuehler, L. Gallman, R. Fluck, G. Zhang, L. R. Brovelli, C. Harder, P. Laporta, U. Keller, *Electron. Lett.* **35**, 567-569 (1999)

12. L. Boivin, M. C. Nuss, W. H. Knox, S. T. Cundiff, *Proceedings of Ultrafast Electronics and Optoelectronics TOPS* **13**, (1997)

13. M. Y. Frankel, J. U. Kang, R. D. Esman, *Electronics Letters* **33**, 2096-2097 (1997)

14. A. V. Krishnamoorthy, D. A. B. Miller, *IEEE J. Sel. Top. Quantum Electron.* **2**, 55-76 (1996)

15. A. Bartels, C. W. Oates, L. Hollberg, S. A. Diddams, *Opt. Lett.* **29**, 1081-1083 (2004)

**Figure Captions:**

**Figure 1** Structure and fabrication process of the Si/Ge-SBR. **a,** Refractive index profile and standing wave pattern (dashed) of the Si/Ge-SBR. **b**, Illustration of the device fabrication, resulting in a reversal of the Si/SiO$_2$ layer sequence seen in reflection.

**Figure 2** Measured properties of the Si/Ge-SBR. **a**, Pump-probe traces of the Si/Ge-SBR taken at various fluence values (solid) along with the cross-correlation of the pump probe laser source (dashed). **b**, Measured and calculated reflectivity of the 6-pair Si/SiO$_2$ Bragg mirror with and without germanium layer.

**Figure 3** Setup and performance of the Erbium-Ytterbium:glass laser. **a**, Schematics of the laser cavity (CM = curved mirror). **b,** Optical spectrum of the Er-Yb:glass laser mode-locked with the Si/Ge-SBR on linear and logarithmic scale. The C-band of optical communications is indicated as the shaded region. **c**, RF-spectrum of the laser and, **d**, background free intensity autocorrelation of the dechirped pulse train.

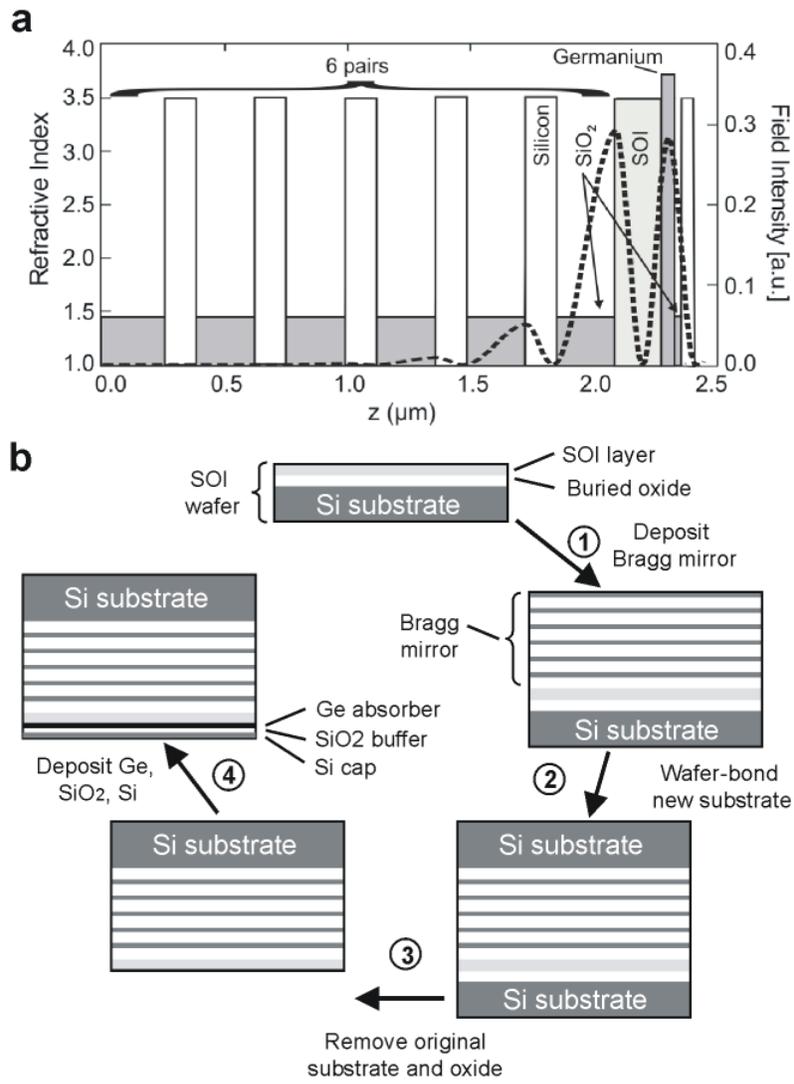

**Figure 1**

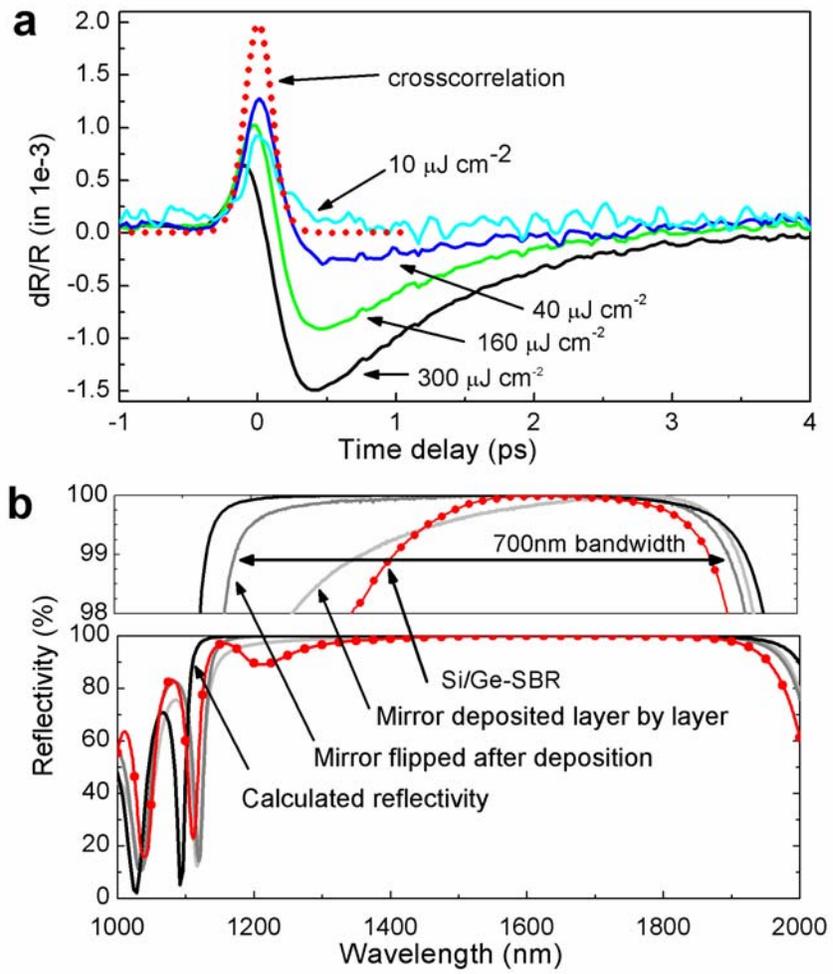

**Figure 2**

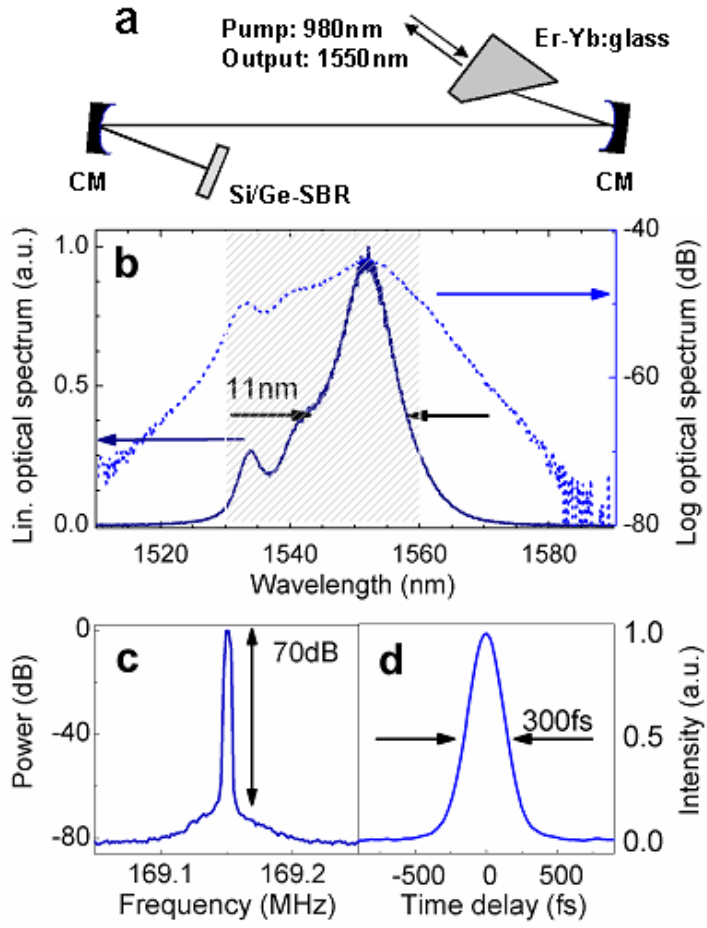

**Figure 3**